# Theoretical Study of High Performance Germanium Nanowire Quantum Dot

Han-Wei Yang, Yung-Feng Wu, Ming-Jung Hsu, Shao-Chen Lee, Ying-Tsan Tang*
Dept of Electrical Engineering, National Central University
No.300, Zhong-da Rd, Zhongli District, Taoyuan City 320317, Taiwan (ROC)
Phone:+886-972-815-512    Email: yttang@ee.ncu.edu.tw

**Abstract**—In this report, we demonstrate that Ge-NWQD (nanowire quantum dots) at low temperatures exhibit apparent Coulomb oscillations than that in Si-NWQD. These oscillations gradually disappear as the temperature increases, indicating the influence of phonon scattering. The increase in Coulomb oscillations enables the device to exhibit multi-level characteristics at low voltage in quantum flash, and the lower barrier high and high mobility of Ge make it advantageous for increasing the storage capacity of quantum flash devices. This research provides design guidelines for optimization of high-performance quantum flash devices.

## INTRODUCTION.

Since the 1980s, research on semiconductor quantum dots (III-V) has been mainly focused on luminescence applications, such as QLED TVs, fluorescent labelling using CdSe, laser printing. These applications have already been commercialized. In the late 1990s, research on quantum computing emerged, including quantum logic gates, quantum key distribution [4], and quantum random walks. [3] The development of quantum information in the past 20 years has matured, mainly based on superconducting quantum dots, which use macroscopic quantum mechanics to verify quantum bits. With the evolution of semiconductor N3 technology, the industry can well control Si processes below 10 nm. Purification techniques for $Si^{28}$ and $Ge^{72}$ quantum dots have accelerated the development of semiconductor quantum dots because purification can dismiss the nuclear spin of the atom, making the spin of electrons or holes on the orbitals more persistent. For example, the coherent time of $Si^{28}$ can reach 50-100 ms at 160K, and that of $Ge^{72}$ can reach 100ms. Ge present higher mobility than Si, allowing faster device operation. Due to the large lattice mismatch between Ge and Si, Ge forms spheres in $SiO_2/Si_3N_4$. The size and position of quantum dots can be manipulated by varying the chemical composition, making the sizes ranging from 5 to 100 nm.[1]

Coulomb blockade oscillations are an important key parameter for quantum computing, providing a direct current-voltage approach to demonstrate quantum interference effects, which reflect the number of states in the system. These discrete quantum states can be used as a unit of non-volatile memory (quantum Flash )[5], where charges can be stored between a gate and a substrate to form storage bits when voltage is applied to the gate. Samsung had adopted the Nano sheet transistors in 2 nm node. Bharadwaj et al.[2] proposed that Si nanowire can show Coulomb diamond behavior through the surrounding gates. In order to verify the advantages of Ge, we compare the physical and electrical differences between Ge-NWQD and Si- NWQD.

## MODEL AND METHOD

Fig.1 shows the simulation model. The gate length of the Ge-NWQD is 3.5 nm.

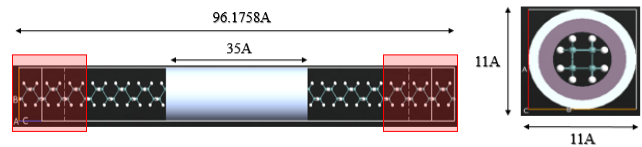

**Fig.1** Schematics of the Ge nanowire device with $L_G$:3.5nm.

Fig.1 shows a (100)-oriented Ge unit cell repeated 1x1x17 times to form a nanowire. To calculate its transport properties, we select three unit cells in each end (marketed by the red box) as the source and drain, where a electron concentration of $1 \times 10^{21}$ $e/cm^3$ were implanted in the source and drain. The isolator with a dielectric constant of 25 were used in this model.

To calculate the current of the Ge nanowire, we used the NEGF method. The transport equation is listed as follows:

$$I(V_L,V_R,T_L,T_R) = \frac{2e}{h} \int T(E)[f(\frac{E-\mu R}{kBT\mathbf{R}}) - f(\frac{E-\mu L}{kBTL})]dE$$

Here, $V_L$ and $V_R$ are the source and drain voltage, $T_L$ and $T_R$ are the temperatures at the two ends, µL and µR are the chemical potentials, f(E) is the Fermi-Dirac function, T(E) is the transmission coefficient, E is the energy, e is the elementary charge, and h is the Planck constant. This formula can calculate the current flowing from the left electrode to the right electrode. To calculate T(E), we use the Meta-GGA method in QATK with the exchange-potential TB09LDA (c=1.088). The cut-off is 30 Ha, and the temperature is set to 10K. The boundary conditions are Neumann in the A and B directions and Dirichlet in the C direction.

## RESULT AND DISCUSSION

Fig.2 shows the Id-Vg curve of the Ge-NWQD at low temperature T=10K under $V_D$: ±5mV. Four current peaks were found in the range of VG: -0.4V to 0.4V. This can be attributed to electron tunneling in and out of the quantum dots. Peaks 1 and 2 correspond to hole state, while peaks 3 and 4 correspond to electron state. As the temperature increases, the Coulomb oscillations will gradually disappear due to the phonon scattering that broadens the electron density of states.

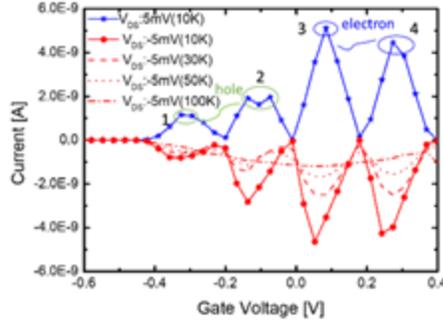

**Fig.2** Transfer characteristics of the $L_G$:3.5nm device at 10 K on linear scale.

Fig.3 plots the $I_D$-$V_D$ curves of Ge-NWQD at T=10K/20K/30K. The curves exhibit step-like features, suggesting the involvement of more quantum states in transport when the transmission windows open. It is implied that Coulomb diamonds can be obtained from Ge-NWQD.

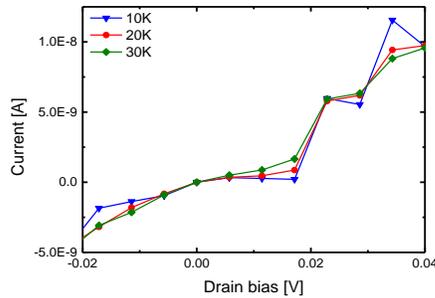

**Fig.3**: Output characteristics of the $L_G$:3.5nm device at 10 K.

Fig4 compares the results of $I_D$-$V_G$ between Si-NWQD and Ge-NWQD. It shows that within the same gate voltage range of -0.2 to 0.4V, Si has two states while Ge has three states, indicating that Ge-NWQD has more storage bits in future memory applications.

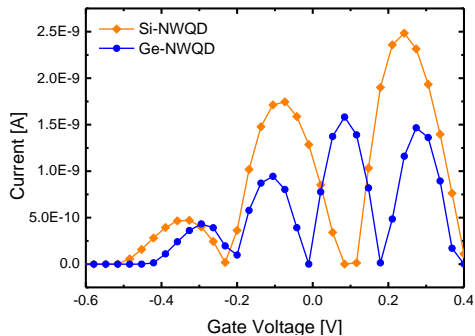

**Fig.4** $I_D$-$V_G$ of the Si-NWQD and Ge-NWQD with $L_G$:3.5nm at 10 K.

Fig.5(a)(b) shows the band diagram of Ge-NWQD. It can be found that the energy levels in the central 3.5nm region shows discrete quantum states. The height of the energy barrier is about 300meV. When increasing the gate voltage, the quantum states will change accordingly, as shown in arrow. We obtained the first and the second at 100meV and 300meV, respectively. Fig.5(c)(d) show the band diagram of Si-NWQD, and it is evident that the spacing of discrete state in Si is larger than that in Ge-NWQD. This is because the lattice constant of Si is smaller than Ge, which also results in a higher energy barrier (450meV) than Si.

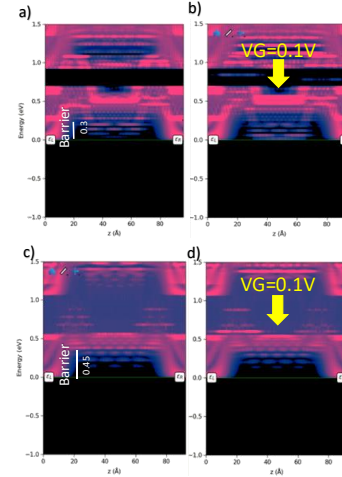

**Fig.5** (a)(b)Band diagram of Ge-NWQD and (c)(d)Si-NWQD.

## CONCLUSION

Our research reveals that Ge-NWQD exhibits more and denser Coulomb oscillations in $I_DV_G$ compared to Si-NWQD. These oscillations gradually disappear with increasing temperature due to phonon scattering. The lower energy barrier of Ge compared to Si also results in faster electron mobility. Within the same operating voltage, The number of Coulomb oscillations in Ge are 1.5 times greater than that in Si, indicated the more storage capacity of bit can be carry out in Ge-NWQD, which is beneficial for future quantum flash memory. These studies provide guidance for the design of high-performance quantum flash memory based on semiconductor devices.

## ACKNOWLEDGEMENTS
We appreciate Prof. Pei-wen Li's discussion and valuable advices. This work is supported by 109-2221-E-008 -093 -MY3,NSTC 112-2119-M-A49-006